\documentstyle[aps,prd,eqsecnum,twocolumn,epsf]{revtex}

\newcommand{\lsim}{\mbox{\raisebox{-1.ex}{$\stackrel{\textstyle<}{\textstyle\sim}$}}}
\newcommand{\square}{\kern1pt\vbox{\hrule height 1.2pt\hbox{\vrule width
1.2pt\hskip 3pt\vbox{\vskip 6pt}\hskip 3pt\vrule width 0.6pt}\hrule height
0.6pt}\kern1pt}
\newcommand{\beq}{\begin{equation}}
\newcommand{\beqn}{\begin{eqnarray}}
\newcommand{\eeq}{\end{equation}}
\newcommand{\eeqn}{\end{eqnarray}}

\begin{document}


\twocolumn[\hsize\textwidth\columnwidth\hsize
\csname@twocolumnfalse\endcsname

\title{Domain Wall Dynamics in Brane World and Non-singular Cosmological Models}

\author{Naoya Okuyama$^1$ and Kei-ichi Maeda$^{1,2,3}$}

\address{$^1$Department of Physics, Waseda University,  Okubo 3-4-1,
Shinjuku,  Tokyo 169-8555, Japan\\[-1em]~}
\address{$^2$ Advanced Research Institute for Science and  Engineering,
Waseda  University, Shinjuku, Tokyo 169-8555, Japan\\[-1em]~}
\address{$^3$ Waseda Institute for Astrophysics, Waseda  University,
Shinjuku,  Tokyo 169-8555, Japan\\[-1em]~}

\date{\today}

\maketitle

\begin{abstract}
We study brane cosmology as 4D (4-dimensional) domain wall dynamics in 5D bulk spacetime.
For a generic 5D bulk with 3D maximal symmetry, we derive the equation of motion of a domain wall and find that it depends on mass function of the bulk spacetime and the energy-momentum conservation in a domain wall is affected by a lapse function in the bulk.
Especially, for a bulk spacetime with non-trivial lapse function, energy of matter field on the domain wall goes out or comes in from the bulk spacetime.
Applying our result to the case with SU(2) gauge bulk field, we obtain a singularity-free universe in brane world scenario, that is, not only a big bang initial singularity of the brane is avoided but also a singularity in a 5D bulk does not exist.
\end{abstract}
\vskip 1pc
]

\section{Introduction}

In the standard model of cosmology, we usually expect an initial singularity in the beginning of the universe.
It is the so-called big-bang singularity.
In fact, Penrose-Hawking singularity theorems\cite{HE} prove that our universe must start with a singularity.
If a singularity exists, physical quantities diverge, and then classical general relativity will break down there.
We need the so-called ``quantum gravity", which is respected to resolve a singularity problem, however it has not been found.
Although quantum gravity is one of the most important subjects in physics, we are now not only looking for quantum gravity but also trying to construct a unified theory of all fundamental interactions.

As such a unified theory, a string/M theory is proposed, which is expected to include quantum gravitational theory\cite{string,Mtheory}.
It suggests that we are living in a 3-brane in a higher dimensional spacetime\cite{akama,Rubakov}.
In such a brane world picture, particles in the standard model are expected to be confined to 3-dimensional (3D) brane, whereas gravitons propagate in the entire bulk spacetime.
As a result, the extra dimensions can be rather large ($\lsim 0.1$ mm)\cite{brane,brane2}.
An alternative brane model is also proposed\cite{RS}.
It has been found that 4D gravity is recovered on the brane in low energy limit.
Based on such a new world picture, a lot of brane world cosmological models have been extensively studied and found new aspects in cosmology\cite{Binetruy,SMS,Kraus,branecosmology}.
Since the spacetime is inhomogeneous even if the 3-brane is maximally symmetric just as a Friedmann-Robertson-Walker spacetime, the analysis will be complicated.
First, the Friedmann equation was obtained by solving 5D inhomogeneous spacetime, and two new effects including dark radiation term were found\cite{Binetruy}.
Then new method to obtain covariantly the effective 4D gravitational equation was found\cite{SMS}.
Another way to study such a brane cosmology is a domain wall dynamics.
The motion of a domain wall in a given spacetime provides the Friedmann equation of a brane world\cite{Kraus}.
It was found that the mass of Schwarzschild-anti de Sitter (AdS) background black hole spacetime corresponds to dark radiation through its tidal force on a domain wall.

Although a brane scenario provides new aspects in cosmology, it does not resolve a singularity problem.
There is a big bang singularity in a simple brane cosmological model based one the Randall-Sundrum model.
Except for the case with no dark radiation, a black hole singularity also exists in the bulk although it is hidden beyond the horizon.

As for a singularity problem, if a string/M theory really gives a unification of all fundamental interactions and particles, it will be free from a singularity.
Then we expect that the effective theory will show such a singularity avoidance.
The effective string/M theory predicts that a bulk space may contain many fields, such as a dilaton scalar field or gauge fields.
In particular, if our 5D bulk spacetime is effectively obtained by dimensional reduction of the 11D Ho\v{r}ava-Witten model\cite{Mtheory}, a dilaton scalar field and U(1) gauge field appear\cite{Lukas}.
Also in an effective action approach of a string theory, the Gauss-Bonnet term as the leading order quantum correction may appear to gravity, in particular, in the case of heterotic string\cite{Gross}.
Also using brane structure, new mechanisms of spontaneous symmetry breaking of gauge field have been proposed\cite{kubo}.
For example, the present standard model, that is SU(3)$\times$SU(2)$\times$U(1) gauge theory, is obtained on the brane in higher dimensional SU(5) gauge theory.
It suggests that higher dimensional bulk contains non-Abelian gauge field in the brane world scenario.
In this viewpoint, a bulk may not be simply Schwarzschild-AdS spacetime, but it would be a various type of geometries by the effects of bulk fields.
Therefore, in this paper we investigate how geometry of a bulk spacetime contributes to dynamics of domain wall.

In the case with U(1) gauge field in a bulk, it is shown that an initial singularity is avoided\cite{Csaki,barcelo,Mukherji}.
However, since the bounce of a domain wall occurs always inside of the inner horizon of Reissner-Nordstrom-AdS black hole, such a bounce may not be realistic because the inner horizon may be unstable and then it will evolve into a singularity\cite{Hovdebo}.
In this paper, we study dynamics of a domain wall in a spacetime containing non-Abelian gauge field in the bulk, and try to resolve a singularity problem.
This is because there is a globally regular spacetime, so-called particle-like solution, as well as non-trivial black holes in the 5D Einstein-SU(2) Yang-Mills system with spherical symmetry\cite{okuyama}.
We adopt SU(2) Yang-Mills field as non-Abelian gauge field.

In the following section, we first derive the equation of motion of 3D domain wall in the 5D bulk spacetime.
In Sec.~\ref{sec:braneuniverse}, after reviewing a brane cosmology in the pure Randall-Sundrum case and the case containing U(1) gauge field, we discuss the model with SU(2) bulk gauge field.
We find a singularity-free brane cosmological model, in which both a bulk singularity and a big bang initial singularity are avoided.
Summary and discussion follow in Sec.~\ref{sec:summary}.
In appendix, we discuss black holes and a particle-like solution of Einstein-Yang-Mills system.

\section{Dynamics of Domain Wall}
\label{sec:domainwall}

The evolution of the universe in brane cosmology is described by the dynamics of 3D domain wall (our universe) moving in a 5D bulk spacetime.
It strongly depends on the spacetime structure of bulk structure.
Hence we investigate the domain wall dynamics in a generic bulk spacetime with 3D maximal symmetry.
We assume that the 3D domain wall ${\cal D}$ is homogeneous and isotropic, and a bulk spacetime is static.
Just as in the Randall-Sundrum model\cite{RS}, we assume that a 5D cosmological constant $\Lambda_5$ is negative.
We introduce a typical scale length $l$ by $l^2=-6\Lambda_5^{-1}$.
Assuming the existence of 3D maximal symmetry, the general form of 5D metric with a negative cosmological constant is written as
\begin{eqnarray}
ds^2=-f_k(r) e^{-2\delta(r)}dt^2+\frac{dr^2}{f_k(r)} + r^2 d\Omega^2_{3,k},
\label{metric}
\end{eqnarray}
where
\begin{eqnarray}
f_k(r)&=&k-\frac{\mu(r)}{r^2}+\frac{r^2}{l^2}.
\label{fdifinition}
\end{eqnarray}
$k$ takes the value of $0$, $1$, or $-1$, corresponding to flat, positive, or negative curvature of 3D maximally symmetric space, which is the same geometry of domain wall ${\cal D}$.
The bulk geometry is represented by two functions, that is a lapse function $\delta(r)$ and a mass function $\mu(r)$.
In this coordinate, a domain wall is located at $r = a(\tau)$.
We assume the $Z_2$-symmetry between the two sides of the domain wall.

The equation of motion of domain wall is given by Israel's junction condition\cite{israel} at the domain wall,
\begin{eqnarray}
K^+_{\mu\nu}-K^-_{\mu\nu}=-8\pi G_5\left(T_{\mu\nu}-\frac{1}{3}Th_{\mu\nu}\right),
\label{junction}
\end{eqnarray}
where $h_{\mu\nu}=g_{\mu\nu}-n_\mu n_\nu$ is the induced metric of the hypersurface ${\cal D}$ and $K_{\mu\nu}=h^\sigma_\mu h^\lambda_\nu\nabla_\sigma n_\lambda$ is the extrinsic curvature of ${\cal D}$, where $n_\mu$ is unit normal vector of ${\cal D}$.
$T_{\mu\nu}$ is the energy-momentum tensor of 4D matter fields on the domain wall.
In Gaussian normal coordinate, we rewrite the metric (\ref{metric}) as
\begin{eqnarray}
ds^2=d\eta^2+\gamma_{ab}dx^adx^b
\end{eqnarray}
near the domain wall ${\cal D}$, where $\gamma_{ab}$ the induced metric in this coordinate ststems.
The condition (\ref{junction}) becomes
\begin{eqnarray}
K^+_{ab}-K^-_{ab}=-8\pi G_5\left(T_{ab}-\frac{1}{3}T\gamma_{ab}\right),
\label{junction2}
\end{eqnarray}
where
\begin{eqnarray}
K_{ab}=\frac{1}{2}n^\mu\partial_\mu\gamma_{ab}.
\label{extrinsiccurvature}
\end{eqnarray}
Assuming $Z_2$-symmetry ($K^+_{ab}=-K^-_{ab}$), we glue two patches of the bulk spacetime which boundaries are given by a domain wall at $r=a(\tau)$, where $\tau$ is a proper time on the domain wall\cite{Grojean}.
In the given bulk metric (\ref{metric}), the 5-velocity of the domain wall $u^\mu=dx^\mu/d\tau$ and the unit normal vector $n^\mu$ are defined by
\begin{eqnarray}
u^\mu&=&\left(e^{\delta(a)}\frac{\sqrt{f_k(a)+\dot{a}^2}}{f_k(a)}, \dot{a}, 0, 0, 
0\right),\\
n^\mu&=&\left(-e^{\delta(a)}\frac{\dot{a}}{f_k(a)}, -\sqrt{f_k(a)+\dot{a}^2}, 0,
0, 0\right),
\end{eqnarray}
where a dot denotes the derivative with respect to $\tau$.
The minus sign appears because the unit normal vector is taken to be the inward direction.
Then we evaluate the extrinsic curvature from Eq.~(\ref{extrinsiccurvature}).
Their spatial components are given by
\begin{eqnarray}
K^+_{ij} (=-K^-_{ij}) =-\frac{\sqrt{f_k(a)+\dot{a}^2}}{a}\gamma_{ij}.
\label{extrinsicdiscontinuity}
\end{eqnarray}
Generalizing the calculation in \cite{visser}, we find the $\tau\tau$ component as
\begin{eqnarray}
K^+_{\tau\tau} (=-K^-_{\tau\tau}) =e^{\delta(a)}\frac{d}{da}
\left[e^{-\delta(a)}\sqrt{f_k(a)+\dot{a}^2}\right].
\label{extrinsic_cur}
\end{eqnarray}

Now we take the energy-momentum tensor of a domain wall in the conventional form as
\begin{eqnarray}
T_{ab}=\rho u_a u_b+P(\gamma_{ab}+u_a u_b).
\end{eqnarray}
This perfect-fluid form includes normal matter fluid described by $\rho_m$ and $P_m$ as well as a domain-wall tension $\sigma$, i.e.~$\rho=\rho_m+\sigma$ and $P=P_m-\sigma$.
Using Eq.~(\ref{extrinsicdiscontinuity}), spatial part of the junction condition (\ref{junction2}) is reduced to
\begin{eqnarray}
H^2+\frac{f_k(a)}{a^2} =\frac{(8\pi G_5)^2}{36}\rho^2
\label{friedmann},
\end{eqnarray}
where $H=\dot{a}/a$ is the Hubble expansion rate, and the rest gives
\begin{eqnarray}
e^{\delta(a)}\frac{d}{da}\left[e^{-\delta(a)}\sqrt{f_k(a)+\dot{a}^2}\right]&=&-4\pi 
G_5\left(P+{2\over 3}\rho\right).\nonumber\\ &&~\label{friedmann2}
\end{eqnarray}

Eq.~(\ref{friedmann}) with Eq.~(\ref{fdifinition}) gives the equation of the domain wall,
\begin{eqnarray}
H^2+\frac{k}{a^2}=\frac{\Lambda_4}{3}+\frac{8\pi G_4}{3}\rho_m +\frac{4\pi
G_4}{3\sigma}\rho_m^2+\frac{\mu(a)}{a^4},
\label{friedmann_rev}
\end{eqnarray}
where
\begin{eqnarray}
G_4 &\equiv& \frac{4\pi G_5^2\sigma}{3},
\label{G4_G5relation}\\
\Lambda_4&\equiv& \frac{\Lambda_5}{2}+4\pi G_4\sigma.
\label{L4_L5relation}
\end{eqnarray}
Eq.~(\ref{friedmann_rev}) corresponds to the Friedmann equation of our brane universe.
When $\mu(a)$ is a constant, i.e.~the bulk is the Schwarzschild-AdS spacetime, we recover the conventional brane universe model, which contains a quadratic term of energy density and dark radiation\cite{Kraus}.

Here we should stress one point about the energy-momentum conservation law of matter fluid on the brane.
Taking a derivative of Eq.~(\ref{friedmann}) and comparing it with Eq.~(\ref{friedmann2}), we find that
\begin{eqnarray}
\dot{\rho}+3H(P+\rho)=\frac{d\delta(a)}{da}\rho \dot{a}.
\end{eqnarray}
This means that the energy-momentum of the matter on the brane is not conserved unless the lapse function $\delta(a)$ is constant.
We discuss two possibilities in order.\\
(1) Suppose that the tension vanishes and the equation of state is given by $P_m=(\gamma-1)\rho_m$.
Then we find the equation for $\rho_m$ as
\begin{eqnarray}
\dot{\rho_m}+3\gamma H\rho_m=\dot{\delta}\rho_m,
\end{eqnarray}
which is integrated as
\begin{eqnarray}
\rho_m=\rho_0 e^{\delta-\delta_0} \left({a\over a_0}\right)^{-3\gamma},
\end{eqnarray}
where $a_0$ is an initial scale factor, $\delta_0=\delta(a_0)$ and $\rho_0$ is an integration constant, corresponding to the initial energy density.
Since $\rho_m \propto a^{-3\gamma}$ in standard cosmology, the energy in a proper volume will decrease when the universe expands if the
lapse function $\delta (a)$ is a decreasing function with respect to $a$.
However, if $\delta$ converges to constant as $a\rightarrow \infty$, then the energy is asymptotically conserved as $\rho_m\rightarrow C_0 a^{-3\gamma}$.
This case includes the equation of state for vacuum energy ($P_m=-\rho_m$), which does not keep constant in the evolution of the universe.
Then if we tune the parameters in order to find zero cosmological constant initially ($\Lambda_4+8\pi G_4 \rho_0=0$), the present small vacuum energy might be explained by the difference between initial and final values of vacuum energy, although the lapse function must be an increasing function.\\
(2) In the case that the tension does not vanish but is constant, we find
\begin{eqnarray}
\dot{\rho}_m+3\gamma H \rho_m=\dot{\delta}(\rho_m +\sigma),
\end{eqnarray}
the integration gives
\begin{eqnarray}
\rho_m&=& e^{\delta-\delta_0} \left({a\over
a_0}\right)^{-3\gamma}\left(\rho_0+\sigma\int_{a_0}^a da~ 
\delta'(a)e^{-\delta(a)}a^{3\gamma}\right).\nonumber\\ &&~
\label{energydensity}
\end{eqnarray}
The second term proportional to a tension appears further in addition to  non-trivial contribution of $\delta(a)$.
If the 5D bulk is asymptotically AdS spacetime, then $\delta(a)$ converges to $0$ such that energy conservation law is recovered as the universe expands.
But it is just a sufficient condition.
In fact, it turns out in next section that it is realized even in quasi-asymptotically AdS bulk spacetime.
In the next section, taking a concrete model, we shall discuss this term in detail.

\section{Dynamics of brane universe}
\label{sec:braneuniverse}

In order to analyze the motion of a domain wall, we rewrite the Friedmann equation (\ref{friedmann_rev}) in the form of a  particle motion with zero energy in a given potential as
\begin{eqnarray}
{1\over 2} \dot{a}^2 +U(a)=0,
\end{eqnarray}
where the potential $U(a)$ is given by
\begin{eqnarray}
U(a)&=&{1\over 2}\left[k-\frac{\mu(a)}{a^2}-\frac{\Lambda_4}{3}a^2\right.\nonumber\\
&&\left.-\frac{8\pi G_4}{3}\rho_m(a)a^2\left(1+{1\over 2\sigma}\rho_m(a)\right)\right],
\label{potential}
\end{eqnarray}
where $\rho_m(a)$ is given by Eq.~(\ref{energydensity}).
The expansion law is modified by a bulk matter flow into (or from) the brane, in addition to quadratic energy density and `dark radiation'.
In what follows, we discuss the Friedmann equation for several brane models in order.

\subsection{Vacuum Bulk Spacetime}

\begin{figure}
\epsfxsize = 3.2in
\epsffile{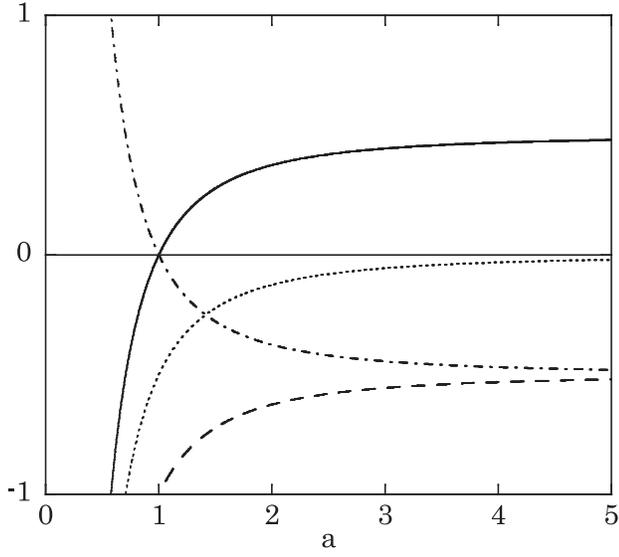}
~\\
\caption{The potential $U(a)$ for the case of a vacuum bulk spacetime.
The solid, dotted, and dashed lines depict $U(a)$ with  $M=1$ for $k=1$, $0$, $-1$, respectively.
We also show the case with  $M=-1$ for $k=-1$ by a dot-dashed line.}
\label{SAdS}
\end{figure}

If bulk spacetime is vacuum just as Randall-Sundrum model, the background is described by Schwarzschild-AdS spacetime\cite{SchBH}, that is $\mu(r)=M$ and $\delta(r)=0$.
Then the potential (\ref{potential}) is
\begin{eqnarray}
U(a) &=& \frac{1}{2}\left[k - \frac{M}{a^2} -\frac{\Lambda_4}{3}a^2\right.\nonumber\\
&&\left.-\frac{8\pi G_4 C_0}{3}a^{(2-3\gamma)} \left(1+{C_0\over
2\sigma}a^{-3\gamma}\right)\right].
\label{friedmann_Sch}
\end{eqnarray}
This case was discussed in detail in\cite{Kraus}.
Unless $M$ is negative, we cannot avoid a big bang singularity.
If we assume that there exists an event horizon (otherwise, unknown effect from naked singularity will come into our brane universe), $M$ is positive for $k=1$ or $0$.
However, if $k=-1$, $M$ could be negative $M>-l^2/4$.
In such a case, there is a chance to avoid a big bang singularity.
In fact, if there is no matter fluid, a big bang singularity is avoided.
However, when matter fluid appears, if $\gamma>2/3$, the potential diverges to $-\infty$ as $r\rightarrow 0$.
Then, a big bang singularity may not be avoided.
We show a schematic diagram in Fig.~\ref{SAdS}.
Furthermore, there is a black hole singularity in 5D bulk spacetime (i.e. at $r=0$).

\subsection{Bulk U(1) Gauge Field Model}

In this case, the 5D bulk spacetime is a Reissner-Nordstrom-AdS spacetime, whose metric is $\mu(r)=M-Q^2/r^2$ and $\delta(r)=0$.
It has a timelike singularity at $r=0$, and two horizons (an event and inner horizons).
Since the lapse function vanishes, the energy-momentum on the brane is conserved.

The potential (\ref{potential}) becomes\cite{Csaki,barcelo,Mukherji}
\begin{eqnarray}
U(a) &=& \frac{1}{2}\left[k - \frac{M}{a^2} + \frac{Q^2}{a^4}  
-\frac{\Lambda_4}{3}a^2\right.\nonumber\\ &&\left.-\frac{8\pi G_4
C_0}{3}a^{(2-3\gamma)}\left(1+{C_0\over 2\sigma}a^{-3\gamma}\right)\right].
\label{friedmann_RN}
\end{eqnarray}
New term corresponding to U(1) charge $Q$ behaves just as stiff matter energy density with the opposite sign.
It works as a repulsive force in the domain wall dynamics.
If there is no matter fluid, a solution turns out to be a catenary type bouncing universe model when $\Lambda_4$ vanishes.
The bounce of domain wall occurs near the inner horizon.
Furthermore, an oscillating universe is realized when $k=1$.
In this model, an initial big bang singularity is avoided.

\begin{figure}
\epsfxsize = 3.2in
\epsffile{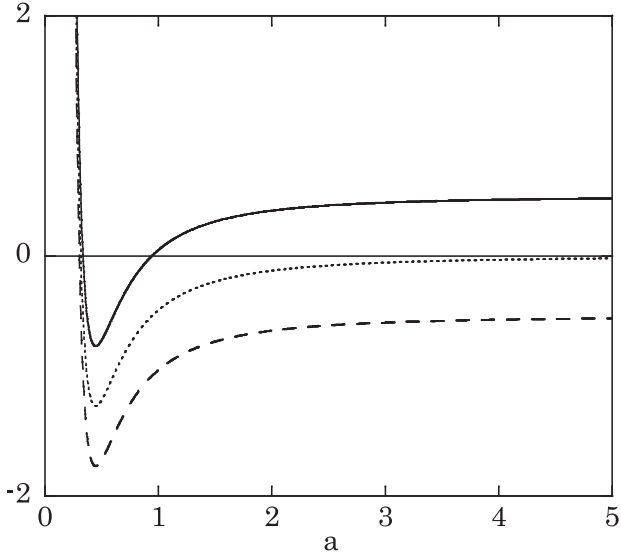}
~\\
\caption{The potential $U(a)$ for the case with a bulk U(1) gauge field.
The solid, dotted, and dashed lines depict $U(a)$ with $M=1$ and $Q^2=0.1$ for $k=1$, $0$, and $-1$, respectively.}
\label{RNAdS}
\end{figure}

However, we can show that any bounce occurs inside the inner horizon, which is expected to be unstable.
In fact, we find that $f(a)>U(a)$ when $\Lambda_4$ vanishes.
The bounce occurs at the inner point of $U(a)=0$, while the horizons exist at the points of $f(a)=0$.
Then we find that the bounce point appears inside of the inner horizon\cite{Csaki,barcelo,Mukherji}.
We show one typical diagram in Fig.~\ref{RNAdS}.
The inner horizon may not appear in the realistic situation and it will evolve into a singularity because of its instability\cite{Hovdebo}.
Hence the bouncing universe is not realized in the case with U(1) charge.

We also have to stress one more important point.
Once we include matter fluid in our model, if $\gamma>1$ including radiation fluid, the potential becomes unbounded from below.
We naturally find a big bang singularity at $r=0$.
Even for the dust fluid $\gamma=1$, in order to obtain a finite potential, there is a constraint for $Q^2$ as
\begin{eqnarray}
Q^2>\frac{4\pi G_4 C_0^2}{3\sigma}.
\end{eqnarray}

\subsection{Bulk SU(2) Gauge Field Model}
\label{cosmology_su2case}

Now we apply our domain wall dynamics to a model with SU(2) bulk gauge field.
The bulk geometry is obtained as a solution of 5D Einstein-Yang-Mills system, whose action is
\begin{eqnarray}
S&=&\frac{1}{16\pi}\int d^5x\sqrt{-g_5}\left[\frac{1}{G_5}(R-2\Lambda_5)
-\frac{1}{g^2}{\rm Tr}({\bf F}_{{\rm YM}}^2)\right],\nonumber\\ &&~\label{action}
\end{eqnarray}
where ${\bf F}_{{\rm YM}}$ is field strength of 5D SU(2) Yang-Mills field and $g$ is its gauge coupling constant.

There are two differences from the models in the previous subsections.
First, there are new types of solutions\cite{okuyama}; a globally regular spacetime, so-called a particle-like solution, and non-trivial black holes, so-called colored black hole solutions, with spherically symmetry ($k=1$).
Those are numerical solutions and have non-trivial lapse function.
We can show that such solutions are stable against spherical linear perturbations if $\Lambda_5$ is negative.
They are 5D version of the solutions formed in the conventional 4D Einstein equations by\cite{BM,bizon,volkov,volkov2,torii,bjoraker,brodbeck,winstanley,Interior1,Interior2}.
In the case of $k=0$, we find that there is no particle-like solution but colored black hole solutions exist.
However, in the case of $k=-1$, there is neither a particle-like solution nor a colored black hole solution.
There is only analytic black hole solution for $k=-1$ (see the detail in appendix).
We should stress that a particle-like solution has no singularity, thus such a bulk spacetime is singularity-free.
Secondly, the obtained solutions are not exactly asymptotically AdS, following the definition by\cite{ashtekar,ashtekar2}.
In this bulk spacetime, some conserved quantities such as the ``mass" diverge.
However, any curvature is finite and the metric approaches AdS as $r\rightarrow \infty$.
So this geometry can be called quasi-asymptotically AdS\cite{okuyama,nucamendi}.
In this quasi-asymptotically AdS spacetime, a lapse function converges to zero as $r\rightarrow \infty$, so the energy conservation law of matter fluid on the domain wall is recovered at large radius.
We shall discuss several bulk models in order.

\subsubsection{A particle-like spacetime}

\begin{figure}
\epsfxsize = 3.2in
\epsffile{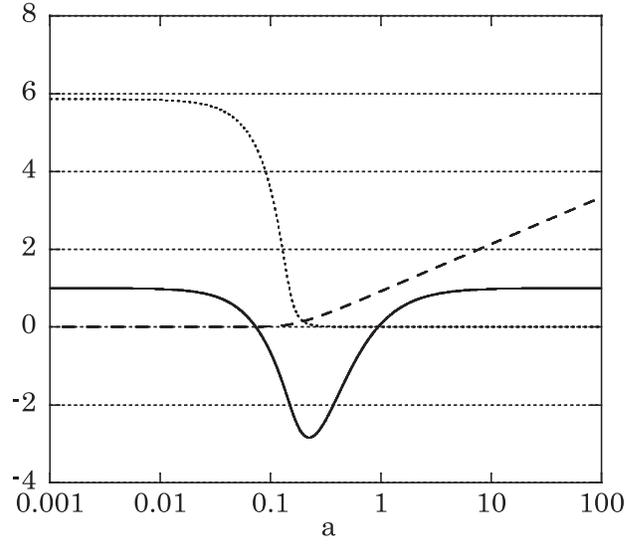}
~\\
\caption{The potential $U(a)$ in a particle-like bulk spacetime.
The solid line depicts the potential $U(a)$.
The dotted and dashed lines depict a lapse function $\delta(a)$ and mass function $\mu(a)$, respectively.
We set the parameters  $l=0.1$ and $b=-7.3$.}
\label{pot_parti}
\end{figure}

First, we consider a brane dynamics in the case that bulk spacetime described by a particle-like solution ($k=1$).
The bulk spacetime is parametrized by $b=w''(r=0)$, where  $w(r)$ is the SU(2) gauge potential.
The regular solution is found for $b_{\rm min}<b<0$.
$b_{\rm min}$ depends on $\Lambda_5$ and is obtained numerically (see Table~\ref{table_b}).
The bulk spacetime is stable against spherical perturbations in the range ($b_{\rm min}\leq$) $b_{\rm s}<b<0$, where $b_{\rm s}$ is the lower bound for existence of a stable solution.
Assuming $\Lambda_4=0$ and $|\sigma|\gg \rho_m$, we find that a domain wall can be located in this bulk spacetime if $b_{\rm min}<b<b_{\rm crit}$.
$b_{\rm crit}$ is maximal value such that $U(a)$ across zero, so no timelike domain wall exists for $b_{\rm crit}<b$.
To find a stable singularity-free universe model ($b_{\rm s}<b_{\rm crit}$), the 5D cosmological constant $\Lambda_5$ is also constrained as $l/\lambda\leq 1.199$ (Table~\ref{table_b}), where $\lambda=(G_5/g^2)^{1/2}$ which is a typical scale length in the Einstein-Yang-Mills system.
Since
\begin{eqnarray}
G_5=\frac{G_4 l}{2},
\end{eqnarray}
which is obtained by Eqs.~(\ref{G4_G5relation}) and (\ref{L4_L5relation}), assuming $\Lambda_4=0$, the restriction for a gauge coupling constant is given by
\begin{eqnarray}
g^2\leq 0.719~\frac{G_4}{l}.
\end{eqnarray}
We depict a typical potential and bulk geometry (metric) in Fig.~\ref{pot_parti}.
Thus, we can realize a singularity-free brane universe in the range $b_{{\rm s}}<b<b_{{\rm crit}}$, i.e.~both big bang singularity and singularity in the 5D bulk spacetime can be avoided.
From the potential in Fig.~\ref{pot_parti}, we easily find that a domain wall dynamics leads to an oscillating universe model.
For the case $\Lambda_4<0$, oscillating solutions are obtained similarly, but the range of $b$ becomes smaller.
For the case $\Lambda_4>0$, catenary type universe solutions are obtained for all particle-like bulk solutions.

\subsubsection{A black hole spacetime}

\begin{figure}
\epsfxsize = 3.2in
\epsffile{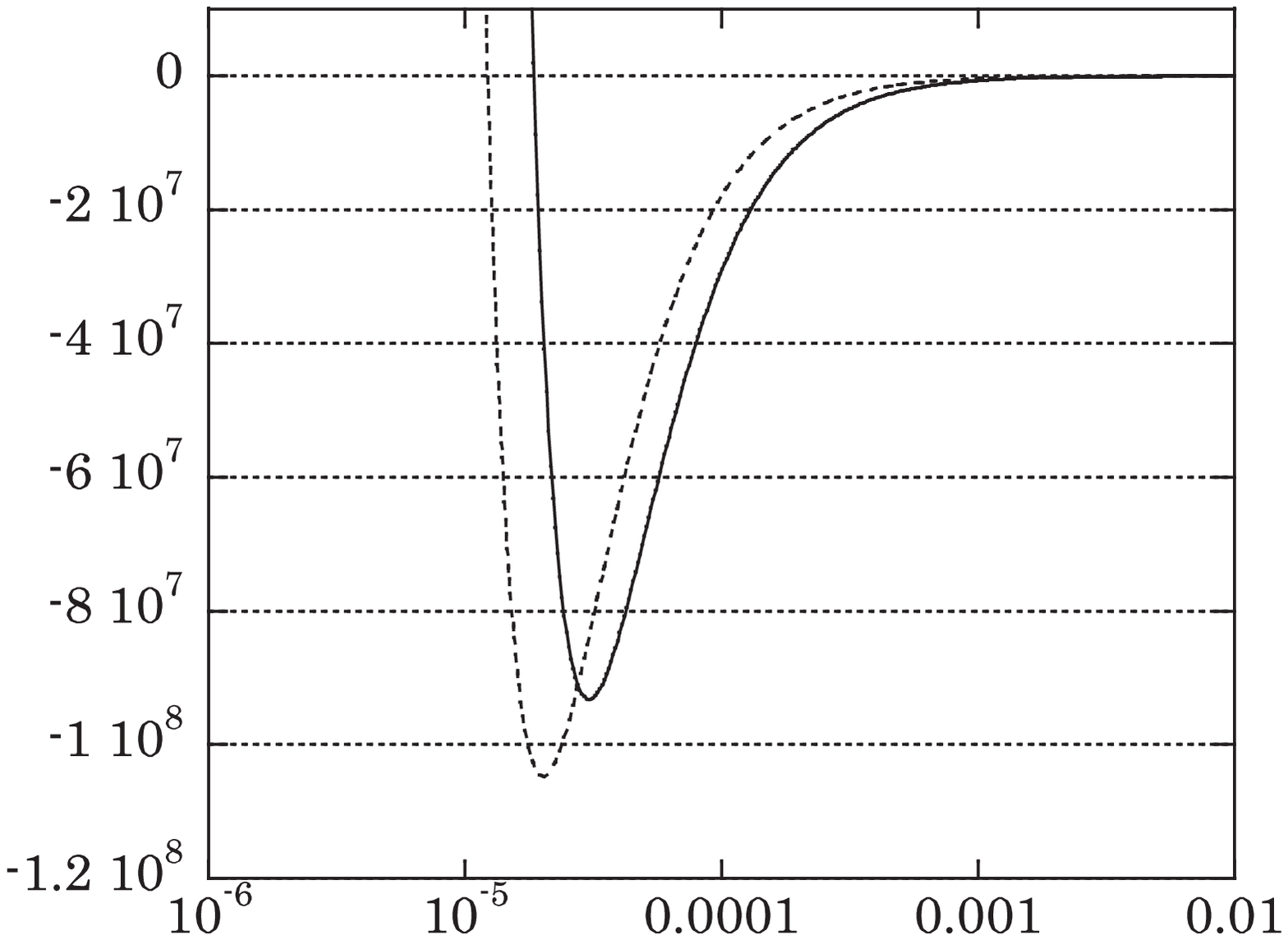}
~\\
\caption{The potentials $U(a)$ for a colored black hole spacetime with an innner horizon.
The solid and dotted line depicts the potential for $k=1$, $w_{\rm h}=0.701$ and $k=0$, $w_{\rm h}=0.56$.
We set the parameters as $l=1$, $r_{\rm h}=1$.}
\label{pot_BHII}
\end{figure}

Next, we consider the case of a bulk black hole spacetime.
Since we know the behaviours for trivial black holes with $\delta(r)=0$, we discuss only non-trivial ones, i.e.~colored black holes.
In this case, a geometry of domain wall allows $k=1$ or $0$.
(In the case of $k=-1$, a bulk spacetime can be the analytic black hole solution, so domain wall solutions are similar to those in Reissner-Nordstrom-AdS case.)
Such black hole solutions are classified into two types by existence of an inner horizon.
If an inner horizon exists, we find a bounce solution in domain wall dynamics.
We show a typical potential in Fig.~\ref{pot_BHII} and bulk geometry in Figs.~\ref{M_BHII},\ref{D_BHII}.
In fact, a catenary type universe can be realized for $k=0$, while an oscillating universe can be obtained for $k=1$.
Just as the case of Reissner-Nordstrom-AdS bulk spacetime, the domain wall always enters inside of the event horizon and escapes through the inner horizon to another quasi-asymptotically AdS spacetime.
However, an inner horizon may be unstable, then a singularity will appear on the Cauchy surface.
Such a brane universe is not a good candidate for a singularity-free cosmological model.

If a bulk spacetime does not have any inner horizon, a domain wall must start from an initial singularity.
In this case, because of a bizarre behaviour of interior structure of colored black hole, $U(a)$ have infinite relative maximums at the point on which $\mu(a)$ approaches 0 most, so domain wall expands repeating acceleration and deceleration.
We show a typical potential in Fig.~\ref{pot_BHI} and bulk geometry in Figs.~\ref{M_BHI},\ref{D_BHI}.
In the case of $k=1$, a domain wall expands to a maximum radius $a_{\rm max}$, where $a_{\rm max}$ satisfies $U(a_{\rm max})=0$.
After then, it contracts to a singularity repeating acceleration and deceleration.
In the case of $k=0$, a domain wall continues expanding to infinity.

We summarize the behaviours of a domain wall in Table.~\ref{table_summary}.

\subsubsection{Energy-momentum conservation}

Now we examine the energy conservation law of  matter fluid on the domain wall.
From our numerical analysis of for a particle-like spacetime, we find that a lapse function is almost constant in the stage of decelerating expansion (or accelerating contraction).
Then the energy in this stage is conserved.
In the region of accelerating expansion (or decelerating contraction), a lapse function is not constant, so energy conservation on the domain wall is broken.
In fact, energy on the domain wall escapes to a bulk spacetime as the universe is expanding.
For the colored black hole bulk, similarly, a lapse function is approximately constant outside the event horizon, and the energy-momentum is conserved there.
Inside the event horizon, energy on the domain wall is not conserved and escapes to a bulk as expanding.
Particularly, much energy flows out to bulk at the point which domain wall dynamics changes from acceleration to deceleration (for the case of colored black hole bulk without inner horizon) or at bounce (for one with inner horizon).
It shows that energy out-flow to the bulk plays an essential role for the acceleration of a domain wall.

\subsubsection{Asymptotic behaviour}

At last, we examine asymptotic behaviour of a domain wall as $a\rightarrow \infty$ (or when $a$ is enough large).
The first term of Eq.~(\ref{energydensity}) describes matter fluid in standard Friedmann equation, which is affected by a lapse function.
The second term is induced energy through some interaction between a brane tension $\sigma$ and a bulk spacetime.
Inserting 
\begin{eqnarray}
\delta(a)'=-{2\over a} (w'(a))^2,
\end{eqnarray}
and the asymptotic behaviour of $w$ as $a\rightarrow \infty$, i.e.
\begin{eqnarray}
w'(a)\rightarrow  {K_0\over a^3},
\end{eqnarray}
with a constant $K_0$, which was obtained in the appendix of\cite{okuyama}, we can evaluate the asymptotic behavior of second term as
\begin{eqnarray}
-\frac{2\sigma K_0^2}{3(\gamma-2)}\frac{1}{a^6}
+{C\over a^{3\gamma}} ,
\end{eqnarray}
for $\gamma\ne 2$, where $C$ is  an integration constant.
This first term behaves just as a stiff matter, while the second term in renormalized into the first term in Eq.~(\ref{energydensity}) because $\delta \rightarrow 0$.
For $\gamma=2$, we find
\begin{eqnarray}
-2\sigma K_0^2\frac{\ln (a/a_1)}{a^6},
\end{eqnarray}
which  behaves similar to a stiff matter fluid but with small difference (the existence of $\ln a$ term).
Hence the second term of Eq. (\ref{energydensity}) is not so important.

The `dark radiation' term behaves
\begin{eqnarray}
\frac{\mu(a)}{a^4}\sim\frac{M}{a^4}+(k-w_\infty^2)^2\frac{\ln a}{a^4}.
\label{darkradiation}
\end{eqnarray}
The first term corresponds to the usual dark radiation, while the second term will be more dominant than the first term because of $\ln a$ term, which is still less important than dust matter fluid.

\section{Summary and Discussion}
\label{sec:summary}

Assuming homogeneity and isotropy for 3D domain wall, we investigate the dynamics of domain wall in a 5D bulk spacetime.
The bulk spacetime is described by a lapse function $\delta(r)$ and mass function $\mu(r)$.
From the Israel's junction conditions, we derive two equations describing dynamics of domain wall, one is Friedmann equation which is same form to\cite{Kraus} replacing the mass of Schwarzschild-AdS black hole by $\mu(r)$.
Another is the modified energy conservation law of matter fluid on the domain wall.
It is worthy of notice that when a lapse function $\delta(r)$ is non-trivial, energy of matter fluid on the domain wall is not conserved.

As for the case with non-trivial lapse function, we study 5D non-trivial black hole spacetime called colored black hole and globally regular spacetime\cite{okuyama}, which are solutions with non-Abelian bulk gauge field.
In this case, energy of the matter fluid on the domain wall leak out to the bulk spacetime by the effect of non-triviality of lapse function as universe expands.
The energy out-flow plays an essential role for the acceleration of a domain wall at the accelerating expanding region.
On the other hand, energy conservation law is asymptotically recovered at large radius of $a$.
While, the mass function works as a repulsive force near $a=0$, so a domain wall solution with appropriate parameters of a bulk spacetime gives an oscillating or a bouncing universe model.
It is worthy of notice that with a globally regular bulk spacetime, a singularity-free stable cosmological model in a brane world scenario can be realized.
Both a big bang singularity and a bulk singularity are avoided.
We summarize our result of domain wall dynamics in Table.~\ref{table_summary}.

In the present analysis, we have studied only SU(2) bulk gauge field.
In the 4D case, however, because we have other gauge symmetries, there are many non-trivial solutions with non-Abelian gauge fields\cite{monopole}.
Then we expect that the similar solutions with different gauge symmetries exist for 5D case as well.
Such a brane picture containing non-Abelian gauge fields in the bulk, through a dimensional reduction of a unified theory, could resolve the singularity problem.
In the effective action of a string theory, we also expect that not only standard forms of gravity (the Einstein-Hilbert action) and Yang-Mills fields exist but also higher-order terms may appear in higher dimensions.
In higher dimensional Einstein-Yang-Mills system with higher order terms of Yang-Mills fields, particle-like solutions with finite energy are found\cite{particle5d}.
It is important to investigate whether in such a string theory or its effective theory, a singularity will not really appear.

\acknowledgements

We would like to thank Takashi Torii, David Langlois and Roy Marrtens for useful discussions and comments.
This work was partially supported by the Grant-in-Aid for Scientific Research Fund of MEXT (Nos.~14047216, 14540281), by a Grant for The 21st Century COE Program (Holistic Research and Education center for Physics Self-organization Systems) at Waseda University, and by the Waseda University Grant for Special Research Projects.

\appendix
\section{Bulk Solutions}
\label{sec:bulk}

In this section, as a non-trivial bulk spacetime, we present static solutions of 5D Einstein-Yang-Mills system with a 3D maximal symmetry, i.e.~a spherical ($k=1$), plane ($k=0$), or hyperbolic ($k=-1$) symmetry.
As for spherically symmetric solutions, we discuss in detail in\cite{okuyama}, where we found a colored black hole  and a particle-like solution.
These correspond  to the Bartnik-McKinnon solution\cite{BM,bjoraker} and a colored black hole\cite{bizon,bjoraker} in 4 dimensions.
Furthermore, an analytic solution of non-trivial black hole is found in 5 dimensions.

In this appendix, we extend our analysis to the case with $k \neq 1$.
We normalize all physical variables by a length scale by $\lambda=(G_5/g^2)^{1/2}$ just as in the previous paper.

We solve the Einstein equations whose action is given by Eq.~(\ref{action}), assuming the metric form by Eq.~(\ref{metric}).
The magnetic part of the vector potential ${\bf A}=A_\mu dx^\mu$ of the SU(2) gauge field having spherically, plane, or hyperbolic symmetry can be obtained by the similar discussion in Appendix A of \cite{okuyama}, and is described by one function $w=w(r)$ as
\begin{eqnarray}
A_t^a&=&0, ~~~ A_r^a=0,~~~A_\psi^a=(0, 0, w),\\
A_\theta^a&=&(wS_k(\psi), -\frac{dS_k(\psi)}{d\psi}, 0),\\
A_\varphi^a&=&(-\frac{dS_k(\psi)}{d\psi}\sin\theta,
-wS_k(\psi)\sin\theta, \cos\theta),
\end{eqnarray}
where
\begin{eqnarray}
S_k(\psi)&=&\left\{
\begin{array}{cc}
\sin\psi & k=1\\
\psi & k=0\\
\sinh\psi & k=-1
\end{array}
\right.,
\end{eqnarray}
and the field strength ${\bf F}=F_{\mu\nu}dx^\mu\wedge dx^\nu$ is written by
\begin{eqnarray}
F_{\mu\nu}=\partial_\mu A_\nu-\partial_\nu A_\mu - [A_\mu,A_\nu].
\end{eqnarray}
Note that pure gauge field (${\bf F}=0$) can be obtained by $w=\pm1$ in the case of $k=1$ and $w=0$ in the case of $k=0$, but there is no pure gauge field in the case of $k=-1$.

With the above ansatz, the Einstein equations and Yang-Mills equation are reduced to a set of equations;
\begin{eqnarray}
\mu'&=&2r\left[f_kw'^2+\frac{(k-w^2)^2}{r^2}\right],\label{equationofm}\\
\delta'&=&-\frac{2}{r}w'^2\label{equationofdelta},
\end{eqnarray}
and
\begin{eqnarray}
\frac{1}{r}(rf_ke^{-\delta}w')'+\frac{2}{r^2}e^{-\delta}w(k-w^2)=0.
\label{equationofw}
\end{eqnarray}
Those with $k=1$ correspond to Eqs. (3.1)-(3.3) of\cite{okuyama}.
In\cite{okuyama}, we discuss a particle-like solution and exterior of a colored black hole in the case of $k=1$.
Here we analyze the other cases ($k=0$ and $-1$) as well as interior of a colored black hole with $k=1$.

First, we search for analytic solutions.
In the case of $k=1$, it was found in \cite{okuyama} that there are two analytic solutions, one is Schwarzschild-AdS spacetime and another is non-trivial black hole spacetime.
For $k=0$, the trivial solution corresponding Schwarzschild-AdS spacetime\cite{SchBH} is obtained as
\begin{eqnarray}
\mu={\cal M}, ~~\delta = 0,~~w= 0,
\end{eqnarray}
while there are no non-trivial analytic solutions.
In the case of $k=-1$, the absence of pure gauge field yields that a magnetic part of SU(2) gauge field cannot provide any trivial solution.
However there is non-trivial analytic solution, as
\begin{eqnarray}
\mu={\cal M}+2\ln r, ~~\delta = 0,~~w=0.
\end{eqnarray}

\begin{figure}
\epsfxsize = 3.2in
\epsffile{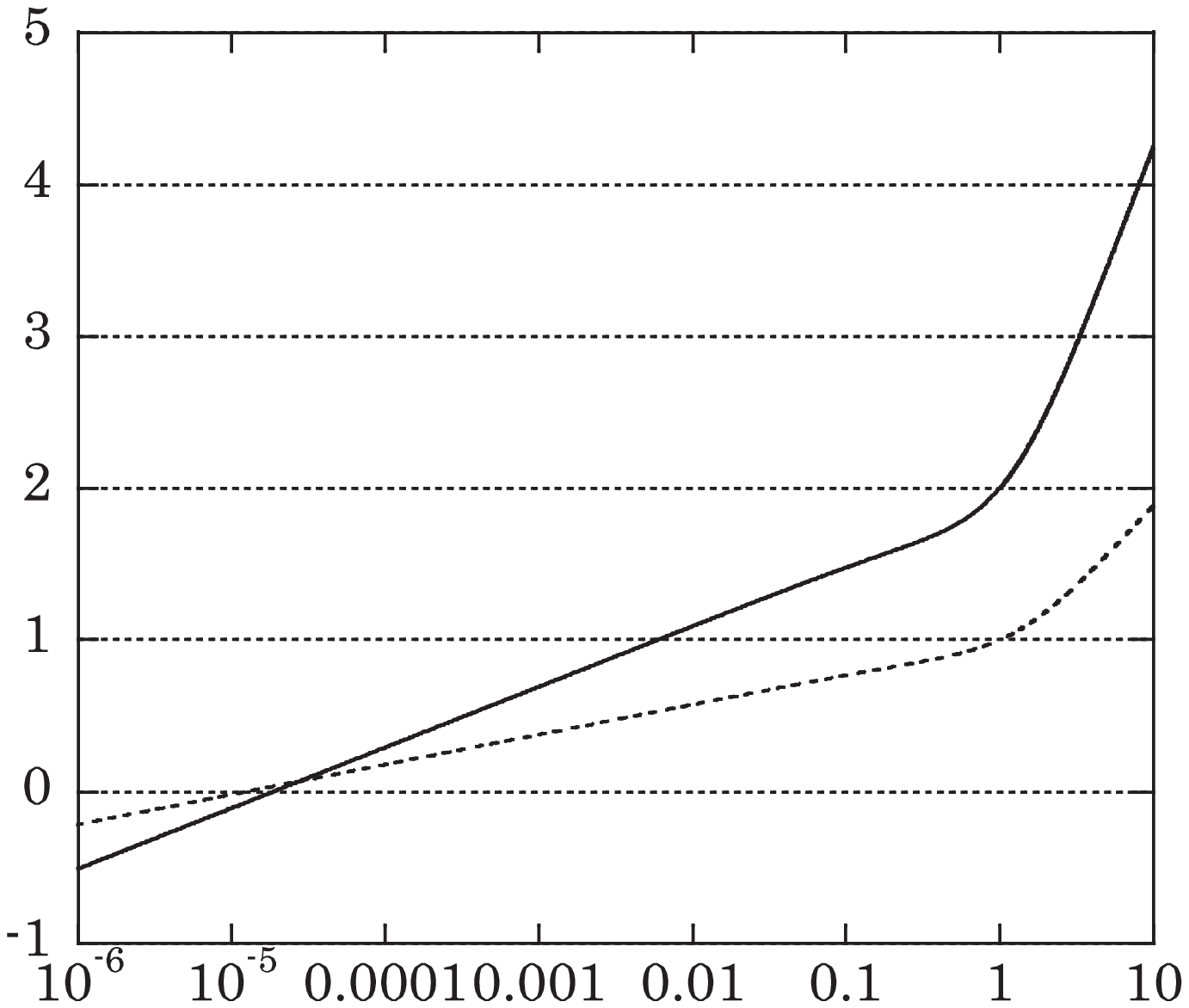}
~\\
\caption{The mass functions $\mu(r)$ for a colored black hole spacetime with an innner horizon.
The solid and dotted line depicts the potential for $k=1$, $w_{\rm h}=0.701$ and $k=0$, $w_{\rm h}=0.56$.
We set the parameters as $l=1$, $r_{\rm h}=1$.}
\label{M_BHII}
\end{figure}

\begin{figure}
\epsfxsize = 3.2in
\epsffile{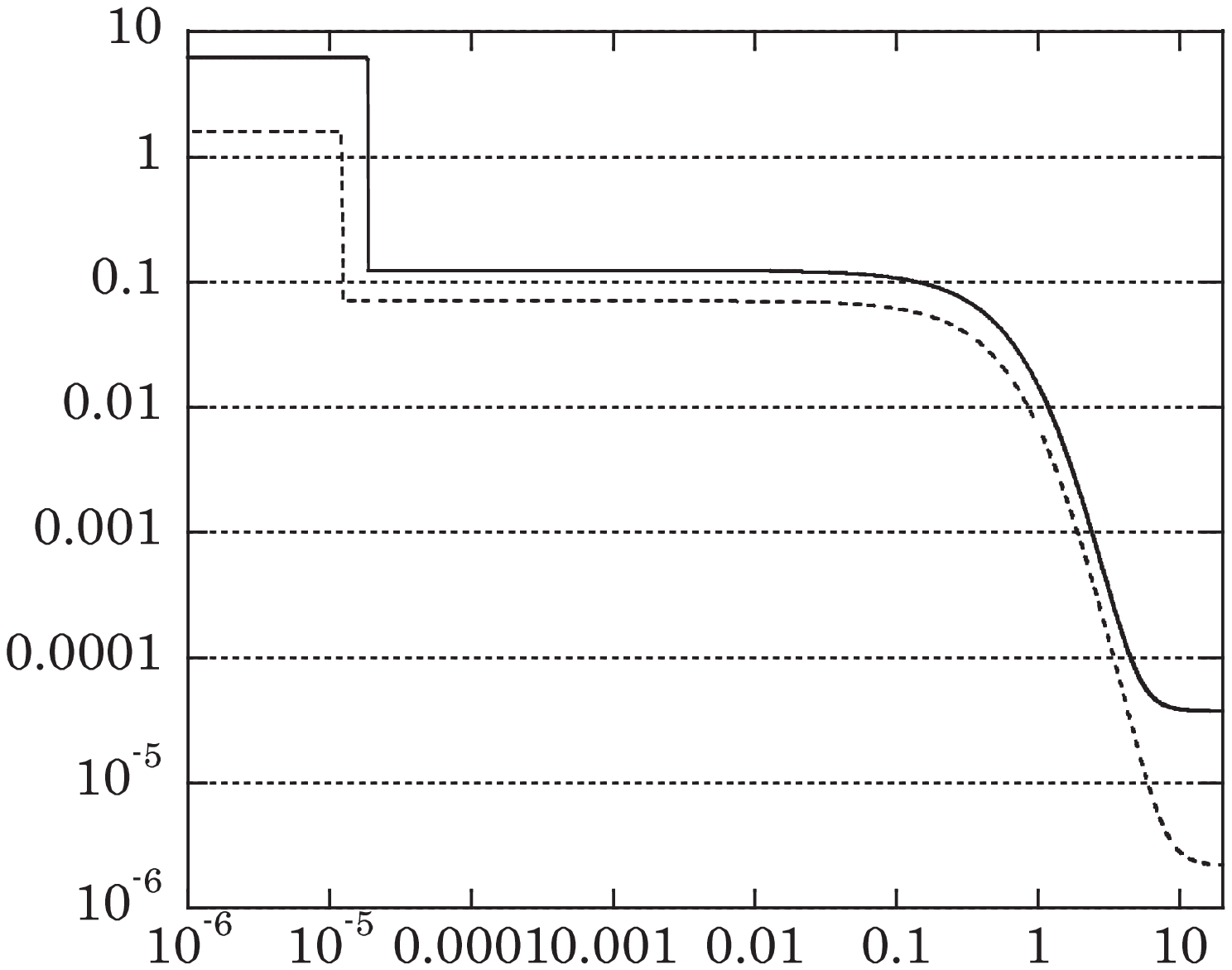}
~\\
\caption{The lapse functions $\delta(r)$ for a colored black hole spacetime with an innner horizon.
The solid and dotted line depicts the potential for $k=1$, $w_{\rm h}=0.701$ and $k=0$, $w_{\rm h}=0.56$.
We set the parameters as $l=1$, $r_{\rm h}=1$.}
\label{D_BHII}
\end{figure}

\begin{figure}
\epsfxsize = 3.2in
\epsffile{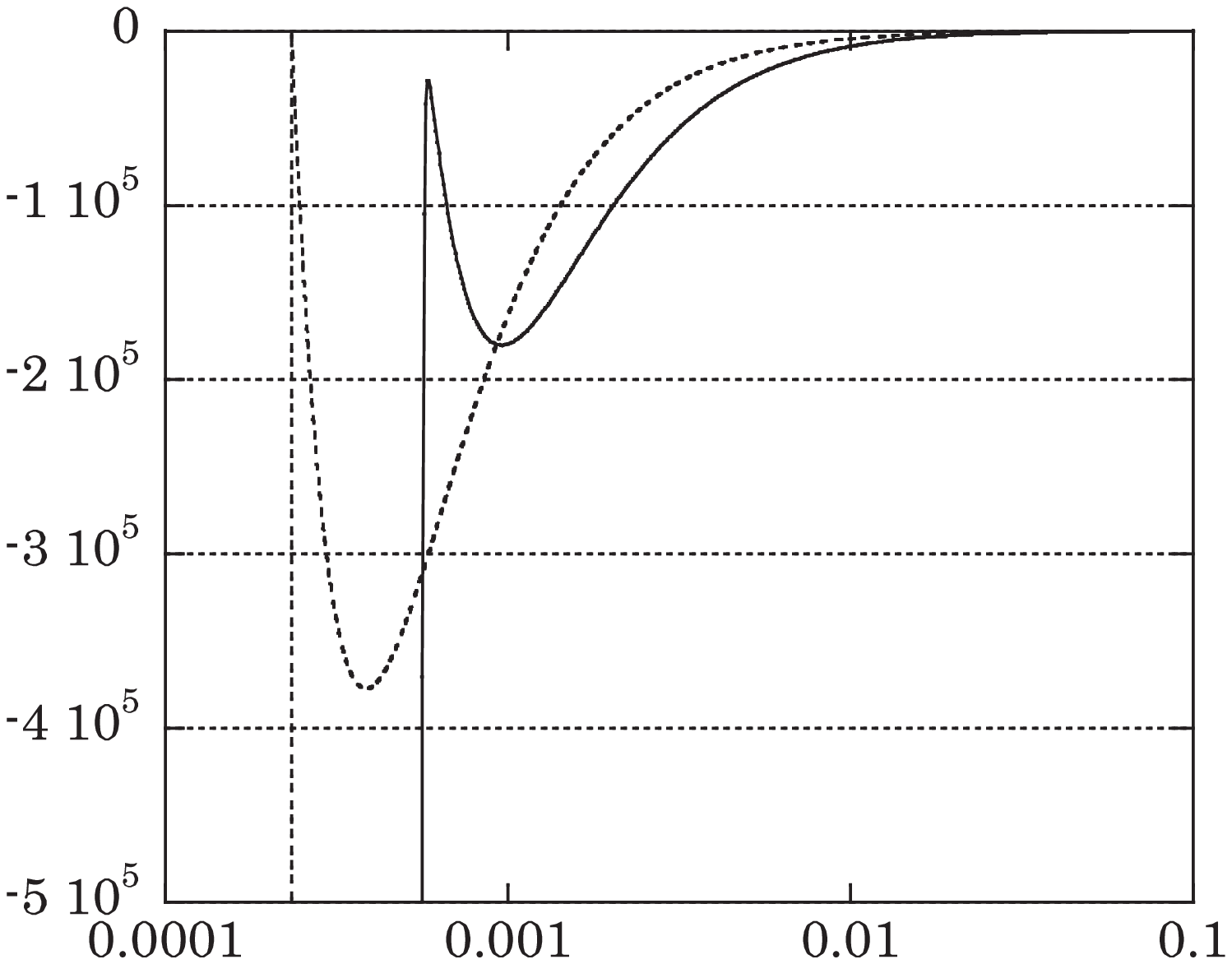}
~\\
\caption{The potentials $U(a)$ for a colored black hole spacetime with no innner horizon.
The solid and dotted line depicts the potential for $k=1$, $w_{\rm h}=0.4$ and $k=0$, $w_{\rm h}=0.7$.
We set the parameters as $l=1$, $r_{\rm h}=1$.}
\label{pot_BHI}
\end{figure}

\begin{figure}
\epsfxsize = 3.2in
\epsffile{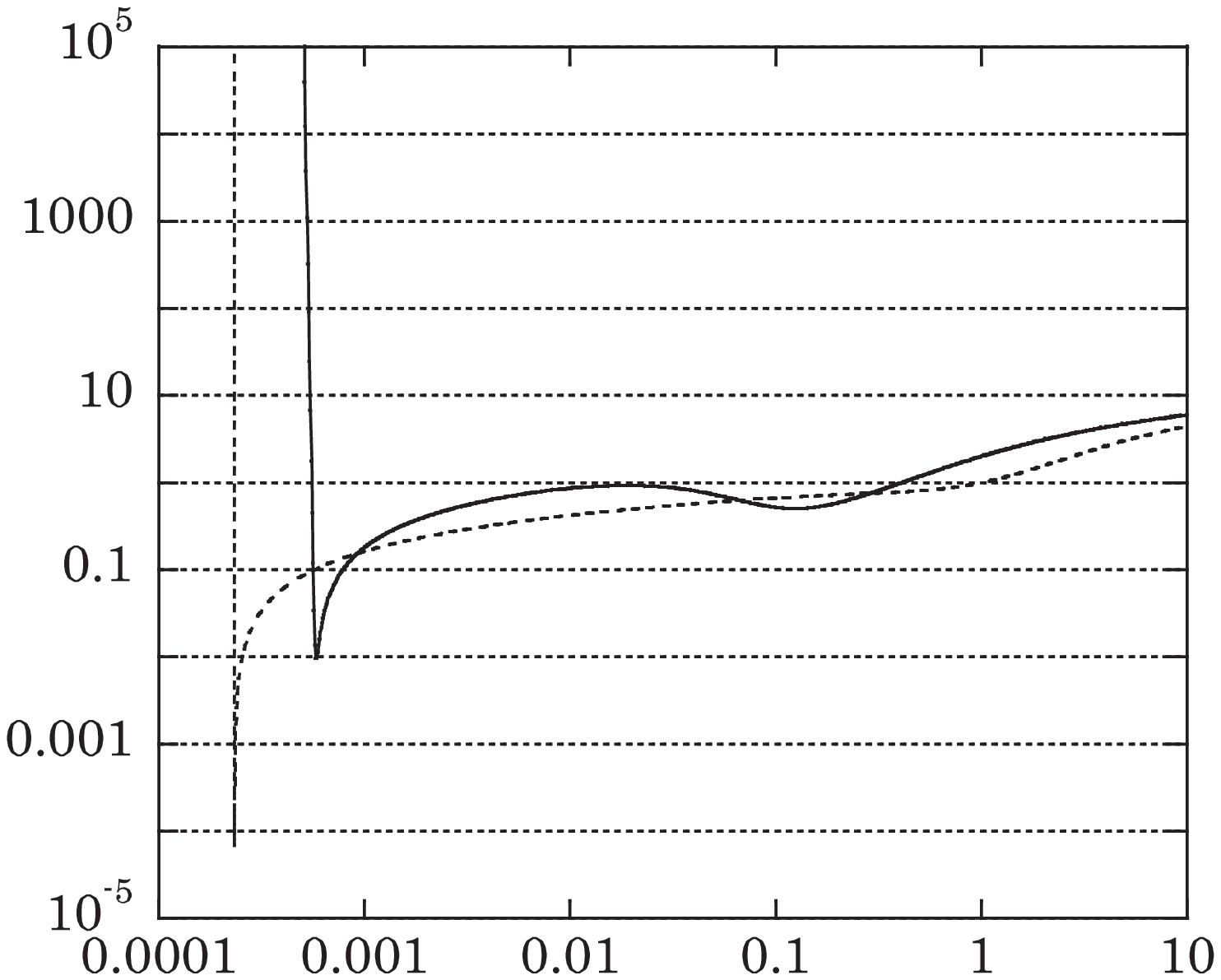}
~\\
\caption{The mass functions $\mu(r)$ for a colored black hole spacetime with no innner horizon.
The solid and dotted line depicts the potential for $k=1$, $w_{\rm h}=0.4$ and $k=0$, $w_{\rm h}=0.7$.
We set the parameters as $l=1$, $r_{\rm h}=1$.}
\label{M_BHI}
\end{figure}

\begin{figure}
\epsfxsize = 3.2in
\epsffile{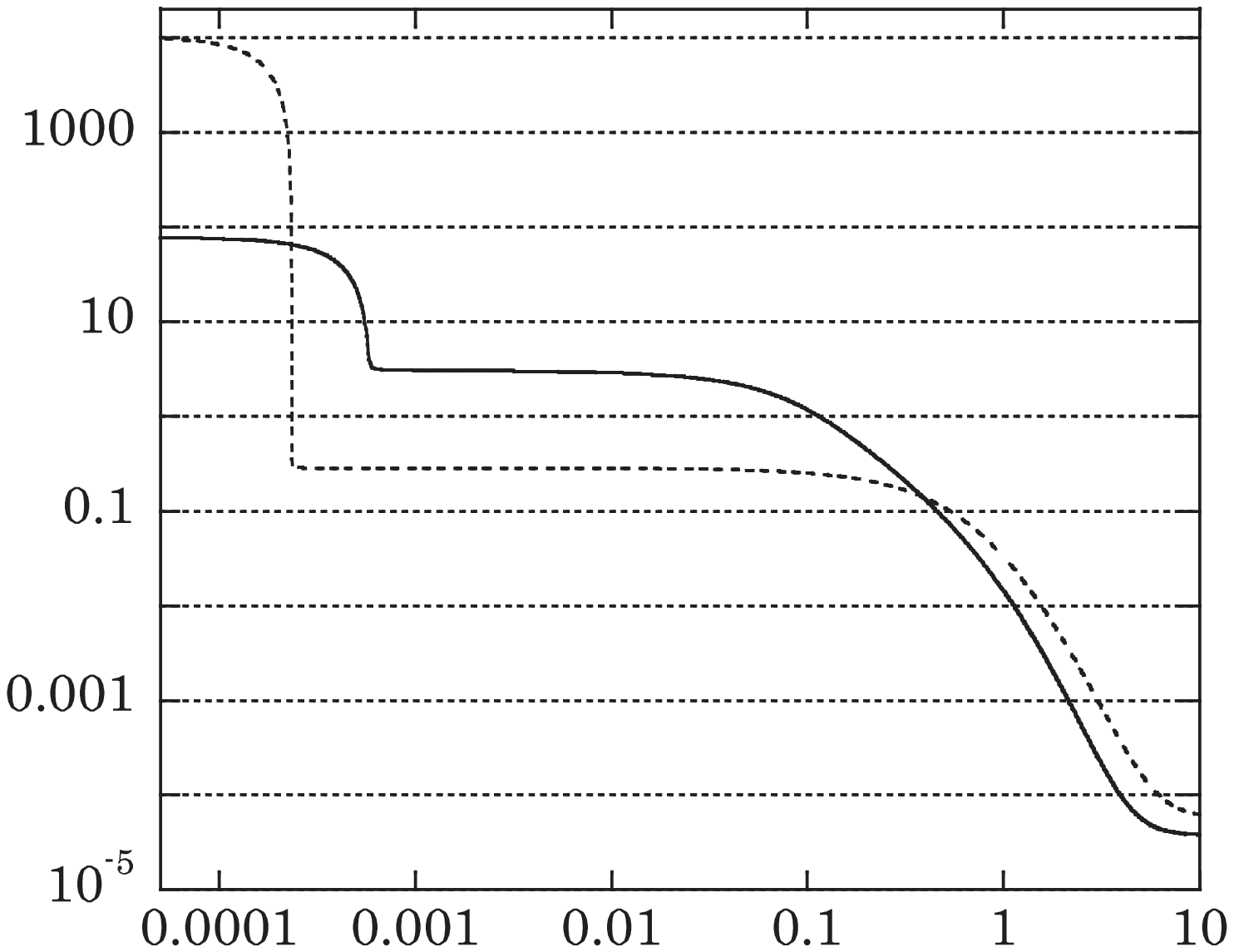}
~\\
\caption{The lapse functions $\delta(r)$ for a colored black hole spacetime with no innner horizon.
The solid and dotted line depicts the potential for $k=1$, $w_{\rm h}=0.4$ and $k=0$, $w_{\rm h}=0.7$.
We set the parameters as $l=1$, $r_{\rm h}=1$.}
\label{D_BHI}
\end{figure}

\begin{figure}
\epsfxsize = 3.2in
\epsffile{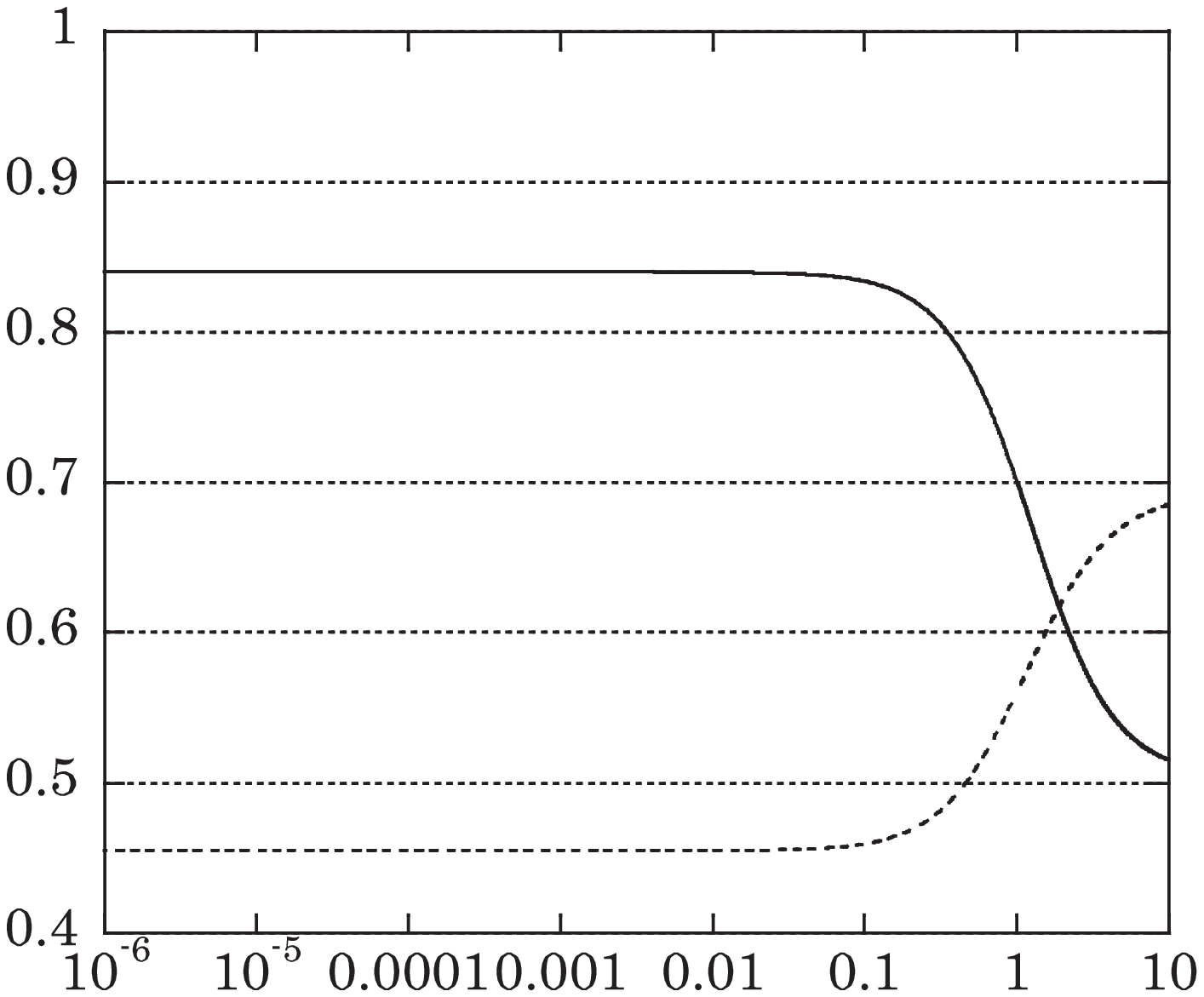}
~\\
\caption{The gauge potential $w(r)$ for a colored black hole spacetime with an innner horizon.
The solid and dotted line depicts the potential for $k=1$, $w_{\rm h}=0.701$ and $k=0$, $w_{\rm h}=0.56$.
We set the parameters as $l=1$, $r_{\rm h}=1$.}
\label{W_BHII}
\end{figure}

\begin{figure}
\epsfxsize = 3.2in
\epsffile{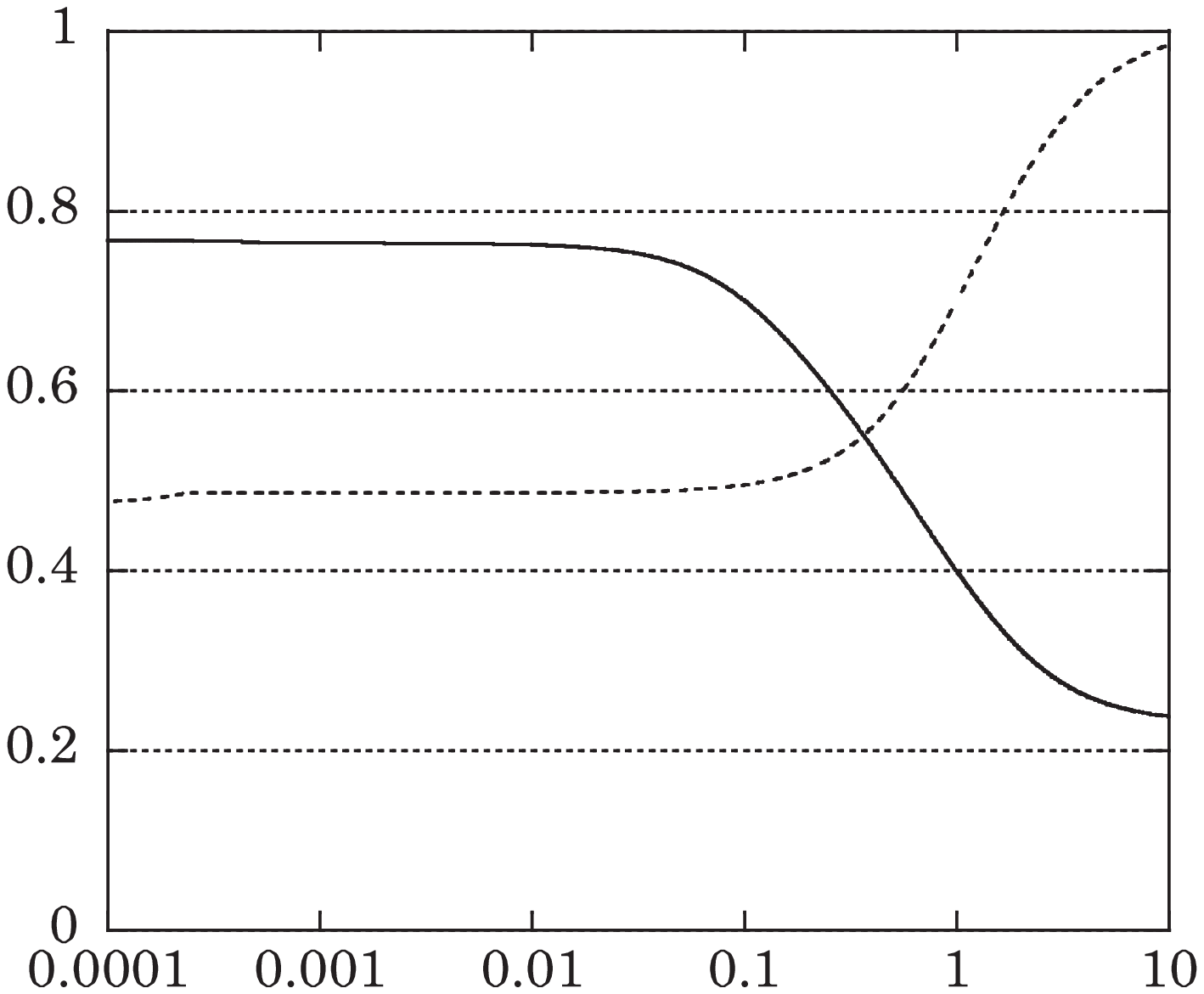}
~\\
\caption{The gauge potential $w(r)$ for a colored black hole spacetime with no innner horizon.
The solid and dotted line depicts the potential for $k=1$, $w_{\rm h}=0.4$ and $k=0$, $w_{\rm h}=0.7$.
We set the parameters as $l=1$, $r_{\rm h}=1$.}
\label{W_BHI}
\end{figure}

Next we search for numerical black hole solutions.
Similarly to the case of $k=1$, there are two free parameters, which are a horizon radius, $r_{\rm h}$, and a value of potential at horizon, $w_{\rm h}=w(r_{\rm h})$.
We find that a black hole solution exists in the parameter range $0<w_{\rm h}<1$ and any value of $r_{\rm h}$ for $k=0$.
However, for $k=-1$ we cannot find any black hole solution with a magnetic part of SU(2) gauge field.
We show our numerical results in Figs.~\ref{M_BHII}-\ref{W_BHI}.
The asymptotic structure at spacelike infinity is similar to that in the case of $k=1$, discussed in \cite{okuyama}.
The colored black hole solutions are classified into two types by existence of an inner horizon.
In 4D, for the solutions without any inner horizon the interior of the colored black hole shows a very bizarre structure\cite{Interior1,Interior2}. 
In 5D, those black holes also show a similar behaviour.
As $r$ decreases below horizon, $\mu$ decreases and approaches zero, but before reaching zero, $\mu$ suddenly grows exponentially and reaches some very large value.
After then, $\mu$ falls down again to very small value which is closer to zero than before.
This behavior is repeated until $r=0$ (see Fig.~\ref{M_BHI}).
For a solution with an inner horizon, $\mu$ across zero and $f_k$ begins to increase to very large value after decreasing to a certain value (see Fig.~\ref{M_BHII}).

Finally we have analyzed a particle-like solution.
For the case of $k=0$ and $k=-1$, there is no solutions which satisfies regularity at the origin. 
Then no particle-like solution exists.
Hence a particle-like solution exists only for the case of $k=1$.

Now we discuss stability of the bulk spacetime obtained above.
For asymptotically flat spacetime in 4D, Volkov et al.~proved that a particle-like solution has $n$ unstable modes if the gauge potential $w(r)$ has $n$ nodes\cite{volkov2}.
It is also true in asymptotically AdS particle-like solutions\cite{bjoraker}.
Since the particle-like solution for asymptotically flat spacetime must have finite node, they are unstable.
On the other hand there is solutions with no node for asymptotically AdS particle-like solutions, so it becomes stable.
Hence, in 5D case we also expect that our bulk spacetime is stable if $w(r)$ has no node.

To analyze the stability, we perturb the metric and gauge potential as
\begin{eqnarray}
\mu(r,t)&=&\mu_0(r)+ \mu_1(r)e^{i\omega t},\\
\delta(r,t)&=&\delta_0(r)+ \delta_1(r)e^{i\omega t},\\
w(r,t)&=&w_0(r)+ w_1(r)e^{i\omega t}\,.
\end{eqnarray}
The  perturbation equations are derived as
\begin{eqnarray}
\mu_1'&=&2 r\left[2f_0w_0'w_1'-{w_0'^2\over r^2}\mu_1-\frac{4(1-w_0^2)w_0}{r^2}w_1\right],
\label{perturbationofm}\\
\mu_1&=&4 rf_0w_0'w_1,
\label{perturbationofm2}\\
\delta_1'&=&-\frac{4}{r}w_0'w_1',\label{perturbationofdelta}
\end{eqnarray}
and
\begin{eqnarray}
&&-\frac{\mu_1}{r^3f_0}(rf_0e^{-\delta_0}w_0')'
-f_0e^{-\delta_0}w_0'\left(\frac{1}{r^2f_0}\mu_1
+\delta_1\right)'\nonumber
\\ &&+\frac{1}{r}(rf_0e^{-\delta_0}w_1')'
+\frac{2e^{-\delta_0}(1-3w_0^2)}{r^2}w_1
=-\omega^2\frac{e^{\delta_0}}{f_0}w_1,
\nonumber \\
~\label{perturbationofw}
\end{eqnarray}
where $\mu_0, \delta_0, w_0$ are given by the unperturbed bulk solution and $f_0=1-\mu_0/r^2+r^2/l^2$.
As boundary conditions for a particle-like solution, we impose $w_1=0$ and $w_1'=0$ at the origin $r=0$, because of regularity.
The asymptotic form of Eq.~(\ref{perturbationofw}) are
\begin{eqnarray}
r^2w_1''+3rw_1'=0,
\end{eqnarray}
which is solved as
\begin{eqnarray}
w_1=A+\frac{B}{r^2}\,,
\end{eqnarray}
where $A$ and  $B$ are integration constants.
We impose $A=0$ because of convergence of total energy flux of gauge field.
We solve them as an eigenvalue problem, restricting $\omega^2\in {\bf R}$.
If $\omega^2$ is always positive, the solution is stable.
We show our result in Table.~\ref{table_b}.
We find that if $b_{{\rm s}}<b<0$, $\omega^2$ is always positive, and the solution is stable.
As we expect, the range of $b_{{\rm s}}<b<0$ corresponds to those of solutions with no node.
Hence we may conclude that the solution is always stable if $w_0$ has no node, while it is unstable if $w_0$ has at least one node.
We also find that for $l\leq 0.779\lambda$, there is not any solutions with a node, so all solutions are stable.
We list up our numerical result in Table.~\ref{table_b}.

\newpage

\begin{table}
\samepage
\caption{The parameter range for which a solution exist.
The  bulk solutions are found in the range $b_{{\rm min}}<b<0$.
They are stable if  $b_{{\rm s}}<b<0$.
A domain wall can be placed in the bulk in the range $b_{{\rm min}}<b<b_{{\rm crit}}$.
We then find that a domain wall can be placed in a stable bulk spacetime if $l\leq 1.199\lambda$.
All bulk solutions are stable for $l\leq 0.779\lambda$.}
~\\
\begin{tabular}{rrrr}
$l/\lambda$ & $b_{\rm min}$ & $b_{\rm s}$ & $b_{\rm crit}$\\
\hline
~~10.0 & ~~-0.644 & ~~-0.013 & ~~-0.633 \\
 5.0 & -0.668 & -0.053 & -0.639 \\
 3.0 & -0.718 & -0.148 & -0.661 \\
 2.0 & -0.801 & -0.329 & -0.707 \\
 1.5 & -0.896 & -0.570 & -0.768 \\
 1.3 & -0.958 & -0.735 & -0.810 \\
 1.2 & -0.998 & -0.837 & -0.837 \\
 1.0 & -1.105 & -1.066 & -0.914 \\
 0.8 & -1.178 & -1.270 & -1.035 \\
 0.7 & -1.390 & $b_{\rm min}$ & -1.125 \\
 0.6 & -1.552 & $b_{\rm min}$ & -1.248 \\
 0.5 & -1.781 & $b_{\rm min}$ & -1.423 \\
 0.4 & -2.128 & $b_{\rm min}$ & -1.691 \\
 0.3 & -2.711 & $b_{\rm min}$ & -2.142 \\
 0.2 & -3.883 & $b_{\rm min}$ & -3.054 \\
 0.1 & -7.411 & $b_{\rm min}$ & -5.808 \\
\end{tabular}
\label{table_b}
\end{table}

\newpage

\begin{table}
\caption{The behaviour of domain wall for several bulk spacetimes.
EH and IH denote an event and an inner horizon, respectively.
BB and BC denote the universe which begins from a big bang singularity, and that ends up with a big crunch singularity, respectively.
Then, BB$\rightarrow$BC means evolutionary path  of the universe.
O denotes an oscillating universe, while C means a catenary type universe.}
~\\
\begin{tabular}{|l||l|l|l|c|}
bulk field & k & bulk spacetime & horizon & evolution \\
\hline
  & $1$ & & & BB$\rightarrow$BC\\
\cline{2-2}\cline{5-5}
vacuum & $0$ & Sch-AdS & EH & BB\\
\cline{2-2}\cline{5-5}
  & $-1$ & & & BB\\
\hline
U(1) or  & $1$ & & & O\\
\cline{2-2}\cline{5-5}
SU(2)  & $0$ & RN-AdS & EH + IH & C\\
\cline{2-2}\cline{5-5}
(electric) & $-1$ & & & C\\
\hline
 &   & analytic BH     & EH + IH & O\\
  & $1$ & colored BH I     & EH      & BB$\rightarrow$BC\\
 &   & colored BH II    & EH + IH & O\\
SU(2)  &   & particle-like & None    & O\\
\cline{2-5}
(magnetic)  & $0$ & colored BH I   & EH      & BB\\
 &   & colored BH II    & EH + IH & C\\
\cline{2-5}
  & $-1$ & analytic BH   & EH + IH & C\\
\end{tabular}
\label{table_summary}
\end{table}

\end{document}